\documentclass[pra,twocolumn,showpacs,amsmath,amssymb]{revtex4}

\usepackage[dvips]{color}
\usepackage{graphicx}

\begin{document}
\title{Comment on ``Quenches in quantum many-body systems: 
One-dimensional Bose-Hubbard model reexamined''}
\date{\today}
\author{Marcos Rigol}
\affiliation{Department of Physics, Georgetown University, Washington, DC 20057, USA}

\begin{abstract}
In a recent paper Roux [Phys. Rev. A 79, 021608(R) (2009)] argued that thermalization in 
a Bose-Hubbard system, after a quantum quench, follows from the approximate Boltzmann distribution 
of the overlap between the initial state and the eigenstates of the final Hamiltonian. 
We show here that the distribution of the overlaps is in general not related to the canonical 
(or microcanonical) distribution and, hence, it cannot explain why thermalization occurs 
in quantum systems. 
\end{abstract}

\maketitle

In Ref.\ \cite{roux09}, Roux argued that thermalization in quantum
systems, after a quench, follows from the approximate Boltzmann 
distribution of the overlap between the initial state and the 
eigenstates of the final Hamiltonian. Roux studied quantum quenches 
in the Bose-Hubbard model and concluded that, for small quenches, 
$|C_\alpha|^2=|\langle \Psi_\alpha |\psi_{ini}\rangle|^2$ 
exhibits an exponential decay typical of a canonical ensemble. 
The distribution of $|C_\alpha|^2$ (or $p_n$, as denoted in \cite{roux09}) 
enters into the computation the infinite-time average $\overline{O}$ of any 
observable $O$, where $\overline{O}=\sum_\alpha |C_\alpha|^2 O_{\alpha\alpha}$,
$O_{\alpha \alpha}= \langle \Psi_\alpha|\hat{O}|\Psi_\alpha\rangle$,
$|\Psi_\alpha\rangle$ are the eigenstates of the final Hamiltonian, and 
$|\psi_{ini}\rangle$ is the initial state. This would explain why 
thermalization was observed in Ref.\ \cite{kollath07} for small quenches. 
Here, we show that the distribution of $|C_\alpha|^2$ is not related 
to the canonical (or microcanonical) distribution and, hence, it 
cannot explain why thermalization occurs in quantum systems.

We study a nonintegrable model of hardcore bosons (HCBs) in a linear 
chain with nearest-neighbor hopping $t$ and interaction $V$, and 
next-nearest-neighbor hopping $t'$ and interaction $V'$. We perform a 
quantum quench from two different initial states that have the same 
energy $E_0=\langle \Psi_{ini}|\hat{H}_{fin}|\Psi_{ini}\rangle$, and 
hence the same effective temperature $T$ \cite{temperature}, in the 
final Hamiltonian $\hat{H}_{fin}$. We utilize full diagonalization 
to study eight HCBs in a 24-site lattice with periodic boundary conditions. 
Translational symmetry is used and the initial state is selected from the 
eigenstates of the initial Hamiltonian with total $k=0$. Further details
about the equilibrium properties and nonequilibrium dynamics of closely 
related HCB systems can be found in Ref.~\cite{rigol09a} and for
spinless fermion systems in Ref.~\cite{rigol09b}.

\begin{figure}[!t]
\centerline{\includegraphics[width=0.477\textwidth]{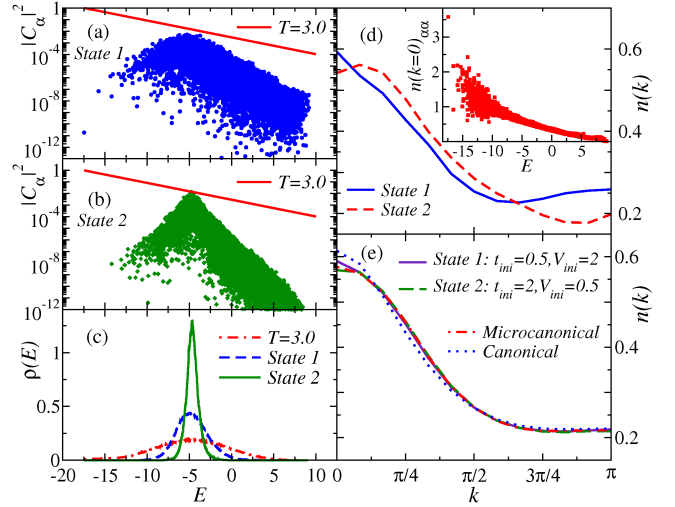}}
\caption{(Color online) Results for a quantum quench
from: $t_{ini}=0.5$, $V_{ini}=2.0$ (State 1), and 
$t_{ini}=2.0$, $V_{ini}=0.5$ (State 2), to 
$t_{fin}=1.0$ (sets the energy scale), $V_{fin}=1.0$. 
In both cases $t'_{ini}=t'_{fin}=0.32$, $V'_{ini}=V'_{fin}=0.32$,
$E_0=-4.62$, and $T=3.0$. (a),(b) $|C_\alpha|^2$ (points) and
canonical (straight line) distributions. (c) Energy distributions 
(d) $n(k)$ of the two initial states. (e) $n(k)$ of the infinite-time average
for both initial states and $n(k)$ of the standard statistical ensembles. 
Inset in (d), $n(k=0)_{\alpha \alpha}$ vs $E$ for all the eigenstates of 
$\hat{H}_{fin}$.}
\vspace{-0.1cm}
\label{fig}
\end{figure}

In Figs.\ \ref{fig}(a) and \ref{fig}(b), we show the distribution of
$|C_\alpha|^2$ for the two initial states selected for our quenches. 
They are compared with the canonical weights corresponding to 
the effective temperature $T$. Two salient features can be seen in those
figures. (i) The distribution of $|C_\alpha|^2$ exhibits a maximum
around the energy $E_0=-4.62$, while in the canonical 
distribution the ground state is always the state with maximal weight.
(ii) The exponent of the exponential decay of $|C_\alpha|^2$ 
is different for the two initial states, i.e., it depends on the 
initial conditions and cannot be predicted by the effective 
temperature of the system. From these results we conclude that, in general, 
the distribution of $|C_\alpha|^2$ is not related to standard statistical 
ensembles and cannot explain thermalization. 

The energy distributions $\rho(E)$,
where $\rho(E)=$ (probability distribution) $\times$ (density of states),
corresponding to both initial states, as well as the one corresponding 
to the canonical ensemble, are shown in Fig.\ \ref{fig}(c). They provide 
guidance to identify from which region of the many-body spectrum 
are the eigenstates that contribute to $\overline{O}$, given the two initial 
states, and to the canonical ensemble result. As expected, 
all the energy distributions peak around $E_0=-4.62$ but they 
are all different from each other. We note that, as discussed in Ref.~\cite{ETH2},
the width of the energy distributions for both initial states is expected
to vanish in the thermodynamic limit, as the width of the canonical 
distribution does.

Figure \ref{fig}(d) depicts the momentum distribution function [$n(k)$] 
of both initial states, which are clearly different from each other. 
In Fig.\ \ref{fig}(e), we present the results of the infinite-time average 
of $n(k)$ for the two initial states, and compare them with the predictions of 
the microcanonical and canonical ensembles. The time averages are virtually 
indistinguishable from each other (independence of the initial conditions) 
and agree with the microcanonical prediction, i.e., thermalization takes place. 
The canonical prediction is slightly different due to finite size effects 
\cite{rigol09a,rigol09b}. Considering that $|C_\alpha|^2$ versus $E$ is different 
for both initial states and from the microcanonical weights, one can 
understand thermalization in terms of the eigenstate thermalization 
hypothesis (ETH) \cite{ETH1,ETH2}. ETH states that the eigenstate expectation 
value of generic few-body observables are very similar between eigenstates 
that are close in energy. From ETH it follows that no matter the weights 
one uses in the average over a narrow window of energies, the result will 
always be the same. The validity of ETH for $n(k)_{\alpha \alpha}$ 
around $E_0=-4.62$ is corroborated by the behavior of 
$n(k=0)_{\alpha \alpha}$ versus $E$ in the inset in Fig.\ \ref{fig}(d). 

\begin{acknowledgments}
We acknowledge support from the Office of Naval Research and from 
Georgetown University.
\end{acknowledgments}


\begin{thebibliography}{99}
\bibitem{temperature} Computed from $E_0$=$Z^{-1}\textrm{Tr}\left\lbrace 
{\hat{H}_{fin}e^{(-\hat{H}_{fin}/{k_B T})}}\right\rbrace$, where 
$Z$=$\textrm{Tr}\left\lbrace e^{(-\hat{H}_{fin}/{k_B T})}\right\rbrace$
and $k_B$ is the Boltzmann constant.
\end{thebibliography}
\end{document}